# Justifying the Adoption and Relevance of Inflation Targeting Framework: A Time-Varying Evidence from Ghana

∗ [a, b]**Nana Kwame Akosah**; [b]**Francis W. Loloh** and [a]**Maurice Omane-Adjepong**

[a] Wits Business School, University of the Witwatersrand, 2 St. Davis Street, Box 5080, Parktown, Johannesburg, South Africa.
[b] Research Department, Bank of Ghana P. O. Box 2674, Accra, Ghana

[*]Corresponding author's email: akosahk@gmail.com; nana.akosah@bog.gov.gh & 1770303@students.wits.ac.za

Abstract

This paper scrutinizes the rationale for the adoption of inflation targeting (IT) by Bank of Ghana in 2002. In this case, we determine the stability or otherwise of the relationship between money supply and inflation in Ghana over the period 1970M1-2016M3 using battery of econometric methods. The empirical results show an unstable link between inflation and monetary growth in Ghana, while the final state coefficient of inflation elasticity to money growth is positive but statistically insignificant. We find that inflation elasticity to monetary growth has continued to decline since the 1970s, showing a waning impact of money growth on inflation in Ghana. Notably, there is also evidence of negative inflation elasticity to monetary growth between 2001 and 2004, lending support to the adoption of IT framework in Ghana in 2002. We emphasized that the unprecedented 31-months of single-digit inflation (June 2010-December 2012), despite the observed inflationary shocks in 2010 and 2012, reinforces the immense contribution of the IT framework in anchoring inflation expectations, with better inflation outcomes and inflation variability in Ghana. The paper therefore recommends the continuous pursuance and strengthening of the IT framework in Ghana, as it embodies a more eclectic approach to policy formulation and implementation.

**I.     Introduction**

The macroeconomic imbalances of the 1970s and early 1980s triggered the adoption of the Economic Recovery Program (ERP) and Structural Adjustment Program (SAP) to help resuscitate the ailing economy. A key component of the reforms was the financial sector reform (FINSAP)[1]. These market oriented reforms led to a more dynamic financial system which encouraged rapid expansion of banking activities (Kovanen, 2011). In addition, the advent of modern payment technology and electronic banking services has resulted not only in the reduction of daily demand for cash but has also supported financial inclusion.

These policy initiatives were envisioned to translate into improved monetary policy transmission mechanism but the general consensus was that fiscal dominance was undermining the effectiveness of monetary policy in Ghana, as persistence fiscal deficits were financed by monetary accommodation. Consequently, strong inertia in inflation expectations became embedded in the economy, underpinned by high inflation and exchange rate volatility. A core challenge for monetary policy, therefore, was to break the inflation inertia, and this was perceived as a prerequisite for achieving the primary objective of price stability. In view of the remaining challenges, a second wave of financial sector reforms was launched in 2001, mainly to address regulatory impediments and its associated corporate governance issues. This reflected the passage of several laws which enhanced the institutional architecture of the financial system. Perhaps, the most significant landmark was the promulgation of the Bank of Ghana Act, 2002, Act (612), which granted the Bank of Ghana operational independence and the establishment of the Monetary Policy Committee (MPC) with the responsibility of formulation and implementation of monetary policy. Upon its inauguration in September 2002, the MPC adopted inflation targeting (IT) as a framework of monetary policy[2]. The ensuing critical policy question therefore is "why did Bank of Ghana abandon the monetary targeting framework and instead adopt the inflation targeting framework?" According to Jahan (2012), the monetary targeting framework works well (i) if the central bank can control the money supply reasonably well and more crucially (ii) if money growth is stably related to inflation over time. Even casual empiricism of the era preceding the adoption of IT revealed that monetary targeting had limited success amid high inflation and inflation volatility accompanied by continual missing of inflation targets, which echoed badly on the credibility of the central bank.

---

[1]This included the liberalization of interest rates and abolition of directed credit, restructuring of financially distressed banks, strengthening of the regulatory and supervisory framework, the establishment of the Ghana Stock Exchange (GSE), and promotion of non-bank financial institutions. The foreign exchange market also witnessed progressive reforms, culminating in the floating of the exchange rate and the establishment of forex bureaus to help eliminate the parallel markets.

[2]While some authors refer to the official announcement to date the adoption of IT (see Kwakye, 2012), we support the position of Bawumia et al. (2008) that even though formal announcement was made in May 2007, Inflation Targeting was launched in the latter part of 2002.

The objective of this paper therefore is to investigate whether at the time Ghana adopted the IT framework, money growth was stably related to inflation. We will do so by employing a simple monetarist state-space model with time-varying coefficients using the kalman filter. To the best of our knowledge, this is the only paper that investigates whether money growth is stably related to inflation in Ghana over time. The extant literature has rather focused on the stability of the demand for money in the context of Ghana, although the results have been inconclusive due to differences in sample periods, data frequencies, monetary aggregates used and model specifications. For instance, Ghartey (1998) found a stable demand function for narrow money in Ghana. Bawumia and Abradu-Otoo (2003) also concluded on a stable long-run relationship between inflation and broad money in Ghana. In contrast, Andoh and Chappell (2002) showed that structural adjustment programs led to structural breaks in the demand for broad money. Similarly, Bawumia, Amoah and Mumuni (2008) asserted that structural reforms and the deregulation of the financial sector have resulted in significant parametric shifts in the demand for broad money in late 1990s and that money growth rates are no longer accurate signposts for forecasting future inflation and real output. Recently, Kovanen (2011) finds a stable money demand function for Ghana but cautions that the stability is not a sufficient condition to conclude that money provides a useful contribution to the inflation dynamics in Ghana. The foregone studies clearly amplify that the current study, which specifically investigate inflation-money growth nexus, is germane as it offers fundamental policy implications for Ghana as well as countries with similar characteristics. Besides, the study can be updated regularly (on monthly basis) as part of the repertoire of tools used by the MPC for its inflation risk assessment.

The rest of the paper is organized as follows. The next section (2) looks at the stylized facts of monetary growth and inflation in Ghana. Section 3 briefly takes up the issues of theoretical framework, methodology and data, Section 4 presents the empirical results and analysis, while Section 5 provides the conclusion.

## 2. Stylized Facts on Ghana

### 2.1 Developments in Monetary Growth and Inflation

For most central banks across the globe, the primary objective of monetary policy is to maintain low and stable prices by using available policy instruments to regulate and control value and quantity of money. This is based on the notion that "inflation is always and everywhere a monetary phenomenon" as professed by famed economists David Hume, Milton Friedman (see section 3). This was premised on the quantity theory of money, derived from the idea that a fast-growing economy allows the government to create more money to help pay for its services without causing inflation. However, inflation results when money growth is faster than real GDP growth.

The historical relationship between money growth and inflation in Ghana is well apparent, especially prior to the year 2002, as depicted in figure 1. In particular, the 1970s and early 1980s were characterised by high and volatile inflation, largely as a result of persistent monetary growth. The continuous monetary expansion was also triggered by incessant monetary accommodation of fiscal expansion fuelled by significant political interference in the operations of central bank (Ayensu, 2007). Monetary situation continued to worsen and inflation was out of control in 1983, culminating in a comprehensive stabilization programme in consultation with the IMF and the World Bank by October 1983. These reforms, inter alia, saw a vigorous promotion of liberal monetary and market-oriented interest rate in the 1988-90, resulting in a revival of interest rate policy as a tool for monetary control. The evolving monetary policy brought some relative stability, as the monetary expansion and excess liquidity which had plagued the economy since the early 1970s, was brought under control. For instance, the growth in money supply declined from 27.3 percent in 1988 to 5.3 percent in 1991. The rate of inflation also plummeted from 31.4 percent in 1988 to 18 percent in 1991 and further to 10 percent at the end of 1992.

Despite the seeming successful implementation of monetary policy under the SAP, the deficit financing through the central bank suddenly resumed, in addition to borrowings from the private sector through the money market in 1993. The increase in money supply during the year was believed to be consequences of a massive injection of cash into the economy to finance electioneering campaigns in the latter part of 1992. The bank intensified its open market operation (OMO) to mop up excess liquidity by introducing a number of BOG bills in the money market. Although this led to reduction in liquidity, it did not succeed in dampening the growth in money supply. By 1994, money supply had increased by more than 50 percent, driven by massive increase in net foreign assets instead of deficit financing. This resulted in restrictive monetary policy stance between 1995 and 1997, reflecting series of hikes in bank and discount rates alongside aggressive OMO operations.

A particular important development was the redefinition of money supply to include foreign currency deposits in 1997 (henceforth called M2+), as part of the restrictive policy measure to fight against excess liquidity and inflation. This further resulted in a new structure of reserve requirement to reflect the inclusion of foreign currency deposits in M2+. These stringent monetary policy measures yielded significant macroeconomic performance by curtailing the rate of money growth alongside declining inflation. For instance, the rate of growth in money supply had dropped from 23 percent in 1997 to 19.2 percent in May 1999 before picking up to 26.8 percent at the end of 1999. At the same time, inflation also dwindled from 20 percent in 1997 (or 47% in 1996) to 9.4 percent in May 1999 before rising to 13.8 percent in December 1999.

However, unfavourable external conditions on the back of falling world prices of cocoa and gold coupled with rising crude oil price during the second half of 1999 through to 2000 disrupted the macroeconomic gains. This reflected monetary expansion, on the back of injection of revaluation gains caused by the significant depreciation of the cedi (by over 49% in 2000). For instance, growth in money supply rose significantly to 46.5 percent in 2000 and moderated gradually to 41.4 percent at the end of 2001. This was mirrored in an astronomical increase in inflation to 40.5 percent at the end of 2000 before easing to 21.3 percent in 2001, calling for further stringent monetary policy measures.

It was against this background that the Bank of Ghana shifted to inflation targeting regime in 2002, following the promulgation of the Bank of Ghana Acts 2002 (Act 612) which granted an operational independence to the central bank and the establishment of monetary policy committee (MPC) as well as strengthening the Bank's primary objective of maintaining price stability. Therefore, the focus of the central bank now shifted from one of considerable fiscal relaxation and monetary accommodation to one of fiscal and monetary stringency. Since 2002, the divergence between money growth and inflation has become more apparent as rising growth in money supply is oftentimes associated with declining inflation and vice versa. This stimulated empirical research on the stability of money demand function, in a quest to ascertain the applicability of the quantity theory of money in the case of Ghana.

**Figure 1: Monetary Growth and Inflation, March 1971 – March 2016**

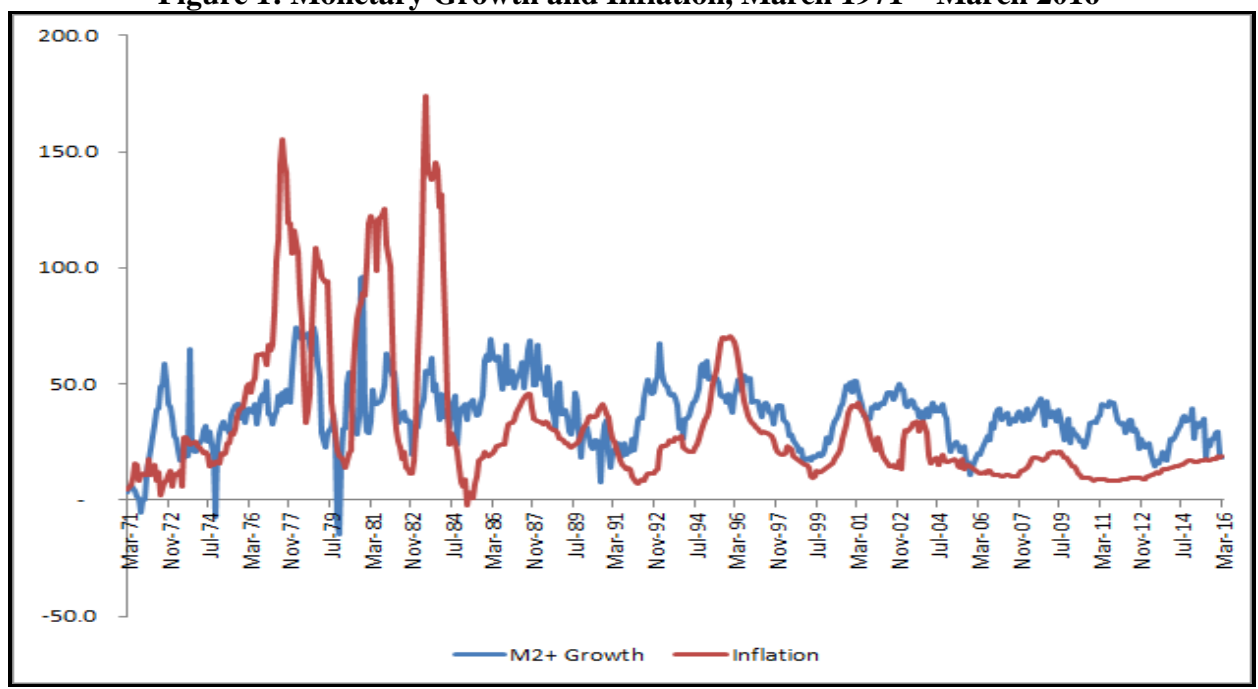

## 2.2    Current IT Framework

The IT framework is based on the notion that policy is designed to target inflation through an inflation forecast. Given that the inflation forecast is a function of many macroeconomic variables, policy reacts to a whole range of variables, not just money supply as in the case of monetary targeting framework. That is, the IT framework is premised on the fact that inflation is not solely a monetary phenomenon, but other factors do influence prices. In this framework, the monetary policy tool of the BOG is the monetary policy rate (MPR), while the operating target is the overnight money market interest rate (Interbank rate). The MPR is the rate around which a policy corridor is defined for central bank's acceptance of deposit and granting of credit to commercial banks. The MPR is set, every other month, at a level considered appropriate by the Monetary Policy Committee (MPC) to deliver the medium-term inflation target. Essentially, the MPR is expected to communicate the stance of monetary policy of the Central Bank and also serve as a guide for all other market interest rates. That is, for the monetary policy to have the desired impact on inflation (the prime objective) and the overall economy, it is critical that changes in MPR affect retail interest rates (e.g. lending and deposit rates) via the wholesale interest rates (preferably, interbank rate) and ultimately influence the overall economy and inflation.

The IT process particularly begins with public announcement of an explicit quantitative target for inflation. Currently, the inflation target for the medium term is 8% (±2%). When inflation stays above target for some obvious reasons, the MPC's aim would be to change interest rates so that inflation can be brought back to target within a reasonable period of time without creating undue instability in the economy.

Monetary Policy Committee (MPC) meets every other month to assess a comprehensive set of forecast based presentations on the real, monetary, fiscal, external, and financial sectors to ascertain medium term risks to the attainment of inflation target. Assessment of deviation of actual inflation from the target determines how the monetary policy rate (MPR) should be positioned by the MPC. Presently, decisions on the MPR are based on consensus at the MPC meetings and communicated to the public via press releases, media encounters, briefing sessions & seminars and publication of Monetary Policy Report. The final part involves the implementation of the policy decision by the relevant departments of the BOG dealing with the repurchase agreements (repo and reverse-repo activities in the interbank markets), BOG and GOG securities markets, bonds issuance and redemption as well as through the sale and purchase of foreign exchange in the interbank market.

In order to increase the efficiency of the transmission of monetary policy and enhancing the transparency and competitiveness of the interbank money market, the BOG has since 2005 introduced a number of policy reforms. Notable among them are the introduction of interest rate corridor, bordered by the reverse repo and repo rates in a bid to influence liquidity in the interbank market; and the amendment of the BOG Act to strengthen the independence of the central bank.

## 3. Theoretical Framework, Methodology and Data issues

### 3.1 Theoretical Framework

In monetary economics, the Friedman's quantity theory of money (QTM) avers that money supply has a direct and proportional relationship with the price level. The QTM was clearly specified as follows:

$$MV_T = P_T Y, \qquad (1)$$

Where $M$ is the total amount of money in circulation on average in an economy during the period, say a year; $V_T$ is the transactions velocity of money, that is the average frequency across all transactions with which a unit of money is spent. This reflects the availability of financial institutions, economic variables, and choices made as to how fast people turn over their money. $P_T$ is the price level associated with transactions for the economy during the period; Y is real index of aggregate transaction (or output).

The theory requires some assumptions to be made about the causal relationships among the four variables in equation (1). There are debates about the extent to which each of these variables is dependent upon the others. Nevertheless, the quantity theory postulates that the primary causal effect is an effect of M on P (that is $\Delta M = \Delta P$).

Even though mainstream economists generally approve that the quantity theory is valid in the long run, there remains contentions about its applicability in the short run. Opponents of the theory contend that velocity of money is unstable and prices are also sticky in the short run. Therefore, the direct relationship between money supply and price level, as stipulated by the quantity of exchange, does not hold over the short run. Consequently, while economists generally accepted the QTM, they placed emphasis on different variable(s) as the main driver(s) of price change. For instance, Marx emphasized on production (output), Marshall, Pigou and Keynes focused on demand for money (thus emphasizing income and demand) instead of money supply using Cambridge equation, while Friedman reiterates the quantity of money.

### 3.2 Methodology

3.2.1 Stability test and Linear state-space model

We employ a battery of procedures to test the stability or otherwise of the relationship between money supply and inflation in Ghana. These include the CUSUM test and recursive coefficients tests conducted to check for parameter (in)stability from a fitted ordinary least square (OLS) model. Once these tests confirm unstable relationship between money supply and inflation, we proceed to model the relationship between the two by employing a simple monetarist state-space framework with time-varying coefficients using the kalman filter.

Our simple monetarist model can be stated as follows:

$$dp\_m_t = \alpha dm\_m_t + \varepsilon_t \qquad (2)$$

Where $dp\_m$ is the demeaned[3] inflation, $dm\_m$ is the demeaned monetary growth, $\alpha$ captures the constant inflation elasticity of money supply growth, $\varepsilon$ is the stochastic error term, and $t$ is time. We examine the fit and stability of the linear relationship between $dp\_m$ and $dm\_m$ by estimating equation (2) using Ordinary Least Squares (OLS).

Should the system in equation (2) fail the stability test and the estimated elasticity of inflation vis-à-vis monetary growth appears to be changing over time, then we will proceed to remodel the relationship in a state space form. Although a general form of a linear state-space model has quite a few components, it can always be written as a system of two equations. In this regard, equation (2) is recast as:

$$dp\_m_t = dm\_m_t \alpha_t + \mu_t \qquad (3)$$
$$\alpha_t = \gamma \alpha_{t-1} + \epsilon_t \qquad (4)$$

Where equation (3) is the measurement equation and equation (4) is the transition equation while $\mu_t \sim i.i.d.(0, \sigma_\mu^2)$ and $\epsilon_t \sim i.i.d.(0, \sigma_\epsilon^2)$. It is obvious from (3) and (4) that the parameter, $\alpha$, is now a time-varying process (i.e., an unobservable state variable) which captures the time-varying inflation elasticity with respect to monetary growth. For simplicity, we assume that the state variable follows a random walk process (i.e., $\gamma = 1$). However, since the disturbance $\mu_t$ in the measurement equation (3) is an error of measurement, the two error terms ($\mu_t$ and $\epsilon_t$) may be contemporaneously correlated. In view of this, the state-space model naturally lends itself to modelling systems with measurement error. There are three unknowns in the system, namely, the state variable ($\alpha_t$), and the two variances ($\sigma_\mu^2$ and $\sigma_\epsilon^2$), which are estimated using the maximum likelihood technique.

There are two main benefits to representing a dynamic system in state space form. First, the state-space form allows unobserved variables to be incorporated into and estimated alongside the observed model. Second, state–space models can be analyzed using the powerful recursive Kalman Filter (KF). The KF due to Kalman (1960, 1963) and Kalman and Bucy (1961) is an algorithm for generating minimum mean square error forecasts in a state-space model. Especially, the KF is a recursive algorithm for sequentially updating the one-step ahead estimate of the state mean and variance given new information. If the Gaussian error is assumed, the KF allows the computation of the log-likelihood function of the state-space model, and in turns, permits the model parameters to be estimated via maximum likelihood methods.

---

[3] Demeaning allows us to estimate the model without a constant

In order to circumvent the possibility of rendering our regression analysis spurious, we examine the stationarity properties of the variables by performing the following Augmented Dickey-Fuller (ADF) test on each variable:

$$\Delta X_t = \alpha + \beta_t(\rho - 1)X_{t-1} + \sum_{j=1}^{p} \rho_j \Delta X_{t-1} + e_t \tag{5}$$

Where $\Delta$ is the first-difference operator, $t$ is a linear time trend, $e$ is a covariance stationary random error term and $\rho$ is determined by Schwartz criterion to ensure serially uncorrelated residuals. The hull hypothesis is that $X_t$ is a nonstationary series and is rejected if $((\rho - 1) < 0$ and statistically significant. The model specification demands that both variables are difference stationary (i.e., I(1)).

### 3.2.3 Dataset

We use monthly data spanning January 1971 and March 2016, a sample size determined not only by the availability and quality of reliable data but also by our desire to capture the dynamic relationship between monetary growth and inflation over the three distinct phases of monetary policy practices in Ghana, namely, direct control, monetary targeting and inflation targeting regimes. CPI (inflation) data is sourced from the Ghana Statistical Service (GSS) while the money supply data is collected from the Bank of Ghana.

## 4. Empirical Results and Analysis
### 4.1 Stationarity Test

The result of the ADF tests is reported in Table 1. It is evident that the null hypothesis of non-stationarity cannot be rejected for the variables at levels. We conclude, therefore, that the variables are non-stationary at levels. Applying the same tests to first differences to determine the order of integration showed that the critical values (in absolute terms) are less than the calculated test statistics for both variables, suggesting that the variables are stationary after first difference. Given that the variables are integrated of the same order, we proceed with the regression analysis, confident that the results we present here are not spurious.

Table 1: ADF Unit Test

| Variables | Levels | p-values | First Difference | p-values |
|---|---|---|---|---|
| Money supply (M2+) | -0.47343 | 0.9845 | -14.14457 | 0.0000* |
| inflation (CPI) | -0.35853 | 0.9888 | -11.34712 | 0.0000* |

Note: *denotes significance at 1%; Critical values are 1% level = -3.975046; 5% level = -3.418117; 10% level = -3.13153. Critical values and one-sided p-values are taken from MacKinnon (1996) and reported in E-views 9.0

## 4.2 Constant Parameter Estimates and Stability Tests

The result from the simple monetarist model (as stated in equation 2) is presented in Table 2. The estimate of $\alpha_t$ (which captures the response of inflation to monetary growth – constant inflation elasticity of monetary growth) is positive and seems to suggest that monetary growth is statistically significant in explaining inflation dynamics in Ghana. However, the model's overall fit seems poor, while further tests suggest model instability. For instance, CUSUM test, which is based on the cumulative sum of the recursive residuals, points to parameter instability as it goes outside the 95% critical bands and only returns into the critical bands after several periods (see Figure 3). Again, the recursive coefficients test, which traces the evolution of $\alpha_t$ as more and more of the sample data are used in the estimation also indicates a significant variation and dramatic jumps, especially prior to 1995, a further evidence of model instability.

A critical look at Figure 4 also reveals that the strong relationship between monetary growth and inflation observed in the early part of the sample has dramatically waned after 1986, with two noticeable downward blips during 1995 and 2002. This strong evidence of parametric shift inherent in the model leads us to conclude that a constant parameter model is inappropriate to capture the true dynamic relationship between monetary growth and inflation developments in Ghana. Therefore, the results contained in Table 1 should be taken with a pinch of salt. In view of this, the appropriate approach to modeling the relationship between monetary growth and inflation is to allow the inflation elasticity of monetary growth to vary overtime.

## Table 2: Constant Coefficient of Inflation Elasticity

| Dependent Variable: DP_M | | | | |
|---|---|---|---|---|
| Method: Least Squares | | | | |
| Sample: 1971M01 2016M03 | | | | |
| Included observations: 543 | | | | |
| Variable | Coefficient | Std. Error | t-Statistic | Prob. |
| DM_M | 0.775278 | 0.073440 | 10.55658 | 0.0000 |
| R-squared | 0.170545 | Mean dependent var | | -4.31E-17 |
| Adjusted R-squared | 0.170545 | S.D. dependent var | | 0.197021 |
| S.E. of regression | 0.179435 | Akaike info criterion | | -0.596165 |
| Sum squared resid | 17.45078 | Schwarz criterion | | -0.588251 |
| Log likelihood | 162.8587 | Hannan-Quinn criter. | | -0.593070 |
| Durbin-Watson stat | 0.103505 | | | |

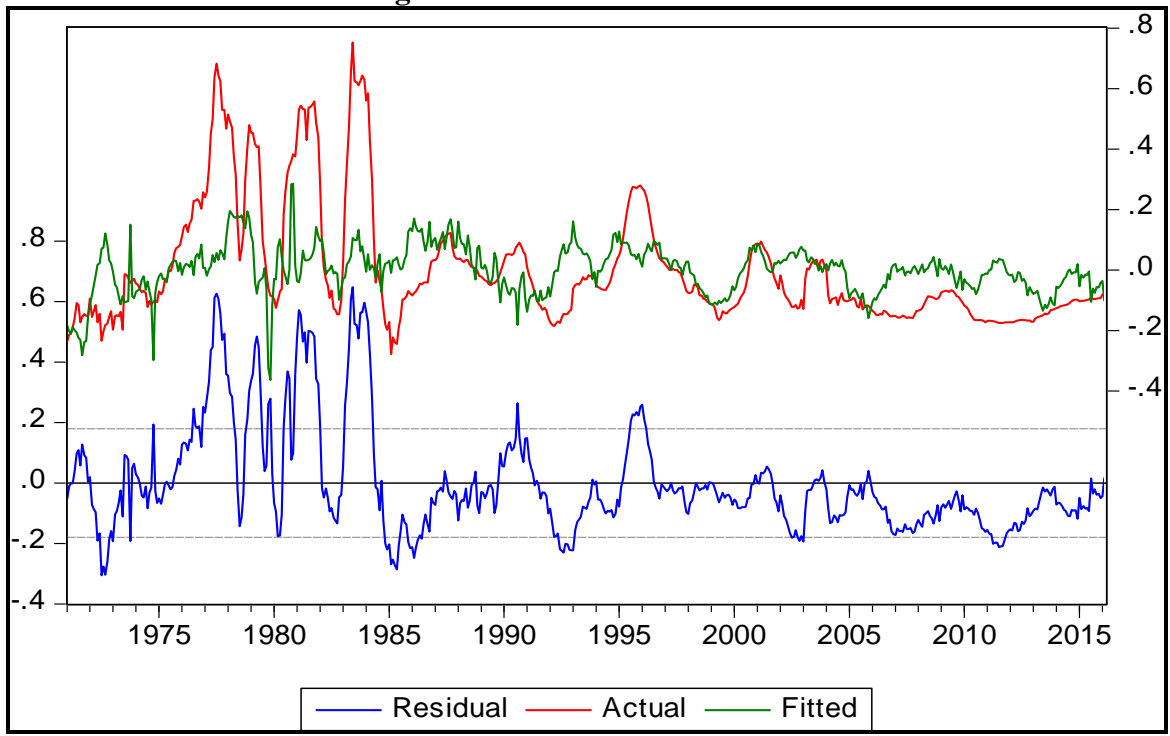

Figure 2: overall fit of the model

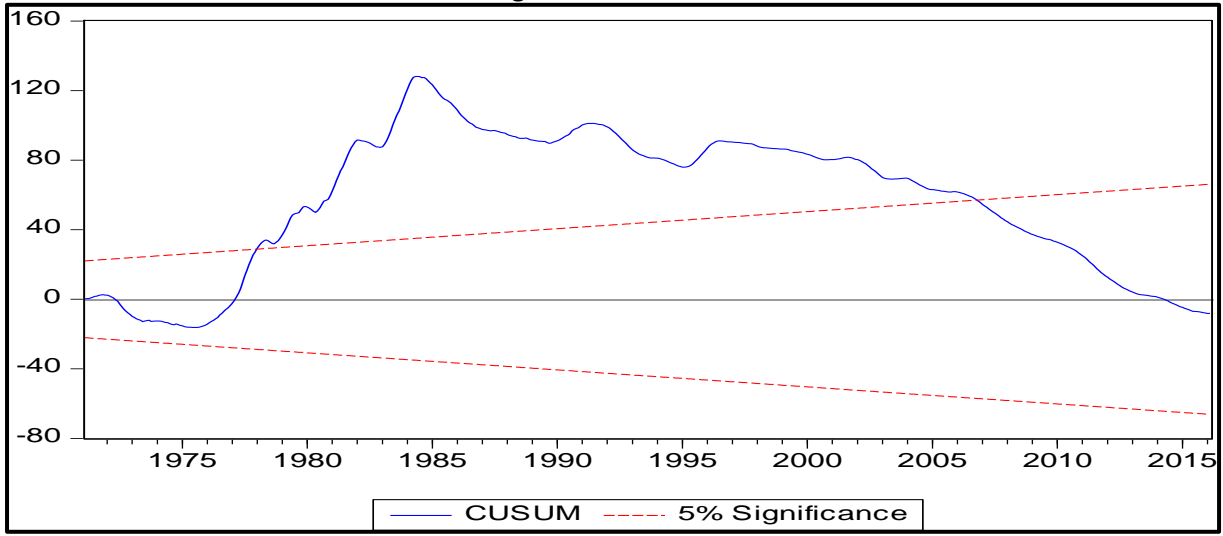
Figure 3: CUSUM Test

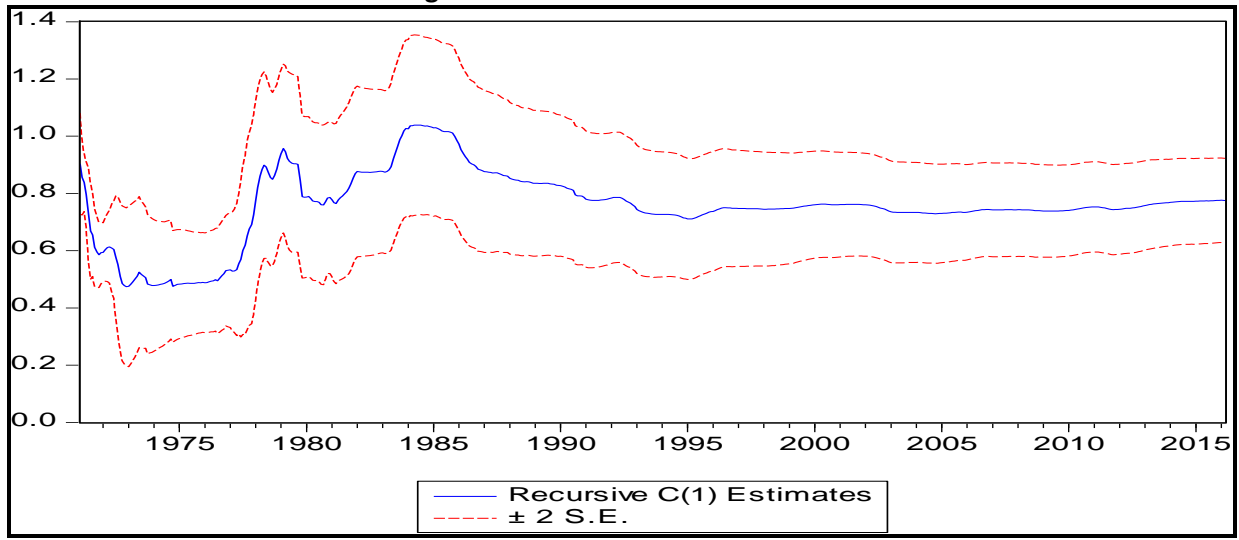
Figure 4: Recursive Coefficients Test

### 4.3 Time-Varying Inflation Elasticity of Monetary Growth (The State Space Model)

Since we have concluded that the system represented by Equation (2) has failed the stability test and the estimated constant inflation elasticity of monetary growth does not reflect the dynamic relationship between inflation and monetary growth, we now proceed to estimate the remodeled relationship in a state space form as depicted in equations (3) and (4). As indicated earlier, $\alpha_t$ (the inflation elasticity of monetary growth) is now a time-varying process. Table 3 presents estimation of equations (2) and (3) using maximum likelihood technique with Marquardt optimization algorithm. The system's convergence was immediately achieved (after zero iteration). The estimated coefficients of $\mu_t$ and $\epsilon_t$, represent the logs of variance of the measurement and transition equations respectively; while $\alpha_t$, is the estimate of the state variable (i.e., the time-varying inflation elasticity with respect to monetary growth).

**Table 3: Time-Varying Coefficient Elasticity of Inflation**

| | | | | |
|---|---|---|---|---|
| Sspace: SSPACE_OUTPUT | | | | |
| Method: **Maximum likelihood (BFGS / Marquardt steps)** | | | | |
| Sample: 1971M01 2016M03 | | | | |
| Included observations: 543 | | | | |
| Convergence achieved after 0 iterations | | | | |
| Coefficient covariance computed using the Huber-White method with observed Hessian | | | | |
| | Coefficient | Std. Error | z-Statistic | Prob. |
| $\mu_t$ | -4.136491 | 0.166651 | -24.82132 | 0.0000 |
| $\epsilon_t$ | -1.025106 | 0.427066 | -2.400344 | 0.0164 |
| | Final State | Root MSE | z-Statistic | Prob. |
| $\alpha_t$ | 0.733380 | 0.891703 | 0.822449 | 0.4108 |
| Log likelihood | 240.3914 | Akaike info criterion | | -0.878053 |
| Parameters | 2 | Schwarz criterion | | -0.862226 |
| Diffuse priors | 1 | Hannan-Quinn criter. | | -0.871865 |

The estimation results reported in Table 3 indicate that $\mu_t$ and $\epsilon_t$ are statistically significant, a confirmation that the system has been estimated more precisely. The estimated coefficients also show that the inflation process, $\mu_t$ is higher than that of the state variable $\epsilon_t$, suggesting that the state variable may be capturing the slow-moving components in the economy (such as real GDP and interest rates) after adjusting for variations in the money supply. Since the coefficients of $\mu_t$ and $\epsilon_t$ are the logs of the variance of the error terms for the measurement and state (transition) equations respectively, we estimate the variance of the measurement equation ($\sigma_\mu^2$) as: ($\sigma_\mu^2$) = exp (-4.136491) = 0.015979, and that of the transition equation ($\sigma_\epsilon^2$) as: $\sigma_\epsilon^2$= exp (-1.025106) = 0.358758.

Table 3 also displays the final one-step ahead[4] value of the state variable, $\alpha_t$, equal to 0.7333 with a corresponding root mean square error (RMSE) value of 0.8917. The estimated coefficient of $\alpha_t$ appears to be high, suggesting that a 1 percent increase in money supply growth will lead to a 0.7333 percent increase in the rate of inflation. However, both the z-statistic (0.822449) and the p-value (0.4108) show that the estimated coefficient is statistically insignificant, implying that money monetary growth exerts a weak positive influence on inflation. This result further suggests that inflation in March 2016 and perhaps immediately after in Ghana was explained by other factors and not necessarily monetary growth. Therefore, monetary targeting would be an inappropriate framework of monetary policy for this era.

But does this outcome necessarily justify the adoption of inflation targeting as a framework of monetary policy by the Bank of Ghana in 2002. A definitive answer demands tracing out the time path of $\alpha_t$, particular around the period in which IT framework was adopted. Figure 5 presents the time path of the inflation elasticity of monetary growth and confirms the earlier evidence of non-constancy of the relationship between inflation and monetary growth. In other words, there is clear evidence that inflation and money supply growth have not been stably related in Ghana.

Despite a brief period of declines in the early 1970s, including a number of negative[5] readings, the inflation elasticity rose generally in much of the 1970s, peaking at 7.2 percent in July 1977. This suggests that a 1 percent increase in money supply leads to a 7.2 percent increase in inflation and is statistically significant[6]. Thereafter, the elasticity broadly trended downward with some negative readings in much of 1979 before ending the decade at 0.25 in December 1979. Average inflation elasticity for the 1970s was 1.44 (see figure 6). Much of this period coincided with the direct control regime of monetary policy and high incidence of fiscal dominance, where large fiscal deficits were predominantly accompanied by monetary accommodation. Without a formalized framework to anchor inflation and inflation expectations, increases in money supply was interpreted by economic agents as engendering inflationary pressures so that as money supply rose, inflation rose even more sharply – in a sense inflation in this decade was everywhere a monetary phenomenon.

---

[4] While this value represents the inflation elasticity of monetary growth for March 2016, the results suggests that the inflation elasticity immediately after March 2016 would not differ substantially from this value (0.7333).
[5] This was a brief period of remarkably low monetary growth and low and stable inflation
[6] Test for coefficient significance is not reported here but is available upon request.

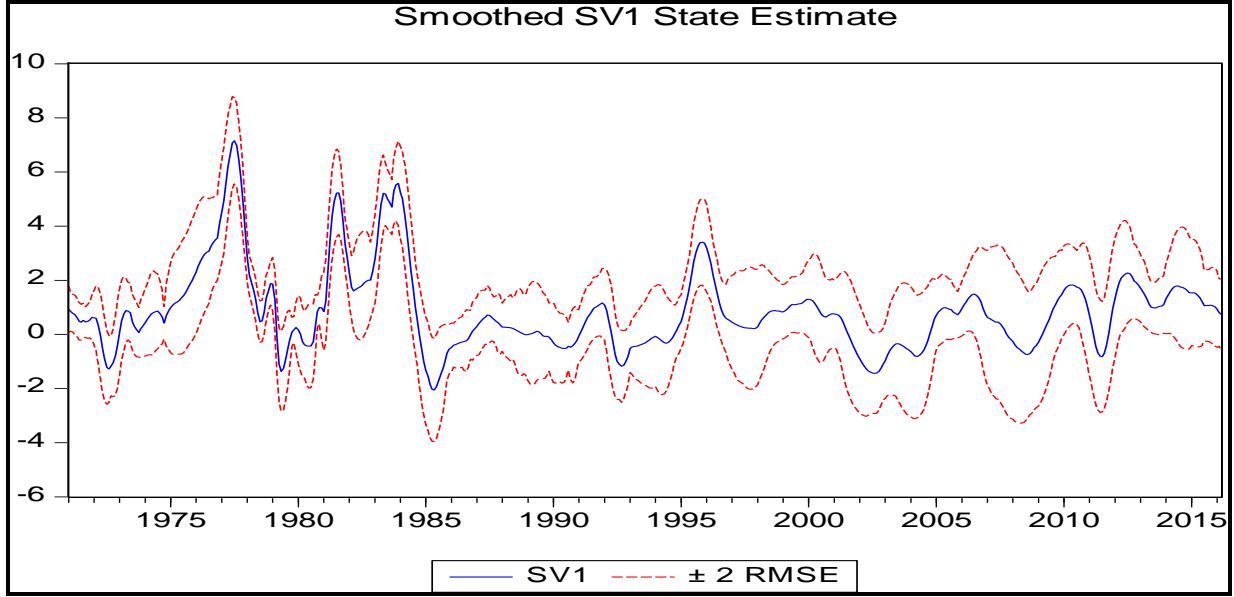

Figure 5: The path of inflation elasticity of monetary growth

Note: SV1 denotes the elasticity of inflation to changes in monetary growth ($\alpha_t$)

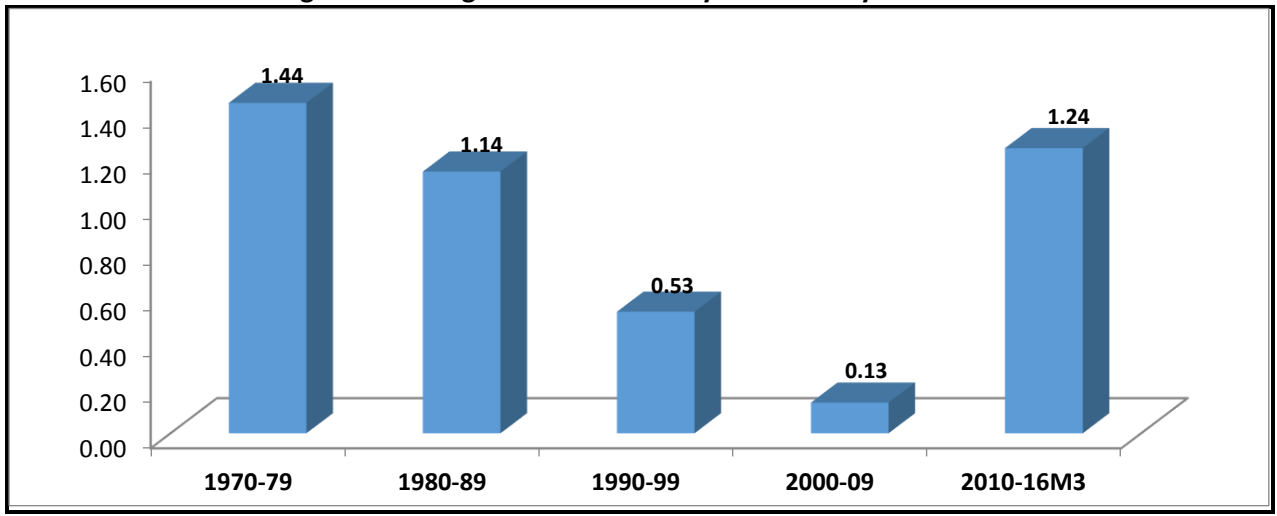

Figure 6: Average Inflation Elasticity of Monetary Growth

Inflation elasticity of monetary growth rose markedly from the early 1980s, reaching a high of 5.6 in December 1983 which was statistically significant. This was largely on account of the persistent monetary growth, triggered by continuous monetary accommodation of fiscal expansion on the back of substantial political interference in the operation of the central bank (Ayensu, 2007). However, following the launch of the comprehensive economic and financial sector reforms from the mid-1980s, the elasticity of inflation declined steadily throughout the remainder of the decade, averaging -0.20 between January 1985 and December 1989. In fact, the average inflation elasticity for the entire decade (i.e. January 1980 to December 1989) stood at 1.14, compared to 1.44 recorded for the preceding decade (see Figure 6).

Interestingly, the inflation elasticity of monetary growth continued its downward trend into the 1990s, registering several negative readings before witnessing some appreciable and statistically significant increases by the middle of the decade. The statistically significant relationship between money supply growth and inflation during this period is a clear justification of the adoption of monetary targeting as a framework for monetary policy by the Bank of Ghana in the 1990s. However, the second half of the decade witnessed broadly weaker and statistically insignificant relationship between money supply growth and inflation. It is essential to note that the average inflation elasticity of monetary growth for the January 1990 and December 1999 period was 0.53, compared with the 1.14 recorded in the preceding decade and 1.44 for the 1970s (see Figure 6). With Bank of Ghana clutching on monetary targeting as a monetary policy framework in the face of lower inflation elasticity and weaker relationship between money supply growth and inflation, inflation outcomes had been less than satisfactory amid high inflation volatility and persistent deviations from targets. Perhaps, the time was ripe to jettison a framework which had promised so much theoretically but delivered little in reality at least in the context of Ghana.

The downward trajectory of the inflation elasticity continued into the 2000s, turning negative between August 2001 and September 2004. Again, the average inflation elasticity for the entire decade, namely, January 2000 to December 2009, stood at 0.13 compared with 0.53 recorded in the 1990s (see Figure 6). This evidence clearly suggests that monetary targeting as a framework of monetary policy at the time would have yielded very little in terms of delivering the desired inflation outcome. And this was precisely the situation at the time where inflation targets were persistently missed, questioning the credibility of the Bank of Ghana and hence leading to entrenched inertia in high inflation expectations. It was against this background that the Monetary Policy Committee (MPC) of the Bank of Ghana upon its inauguration in September 2002 decided to abandon its monetary targeting framework in favour of inflation targeting as a framework of monetary policy.

Nevertheless, the trajectory of the inflation elasticity of monetary growth generally turned upward from 2010, as the average elasticity rose to 1.24 exceeding that of the 1980s, 1990s and 2000s (see Figure 6). The sub-sample analysis, as exhibited in figure 7 and Appendix A1, also shows that inflation elasticity of money supply growth inched up to 2.6 as at end-December 2010 and 2.1 for end-December 2012, and were all statistically significant at 10 percent alpha levels. The strong inflation elasticity of monetary growth in 2010 may be attributed to the implementation of the single spine salary structure (SSSS), resulting in a significant growth in wages and salaries (increase in money supply) and creating strong aggregate demand. Likewise, the fiscal excesses during the presidential and parliamentary election in 2012 also appear to explain the pick-up in inflation elasticity of money growth during the era.

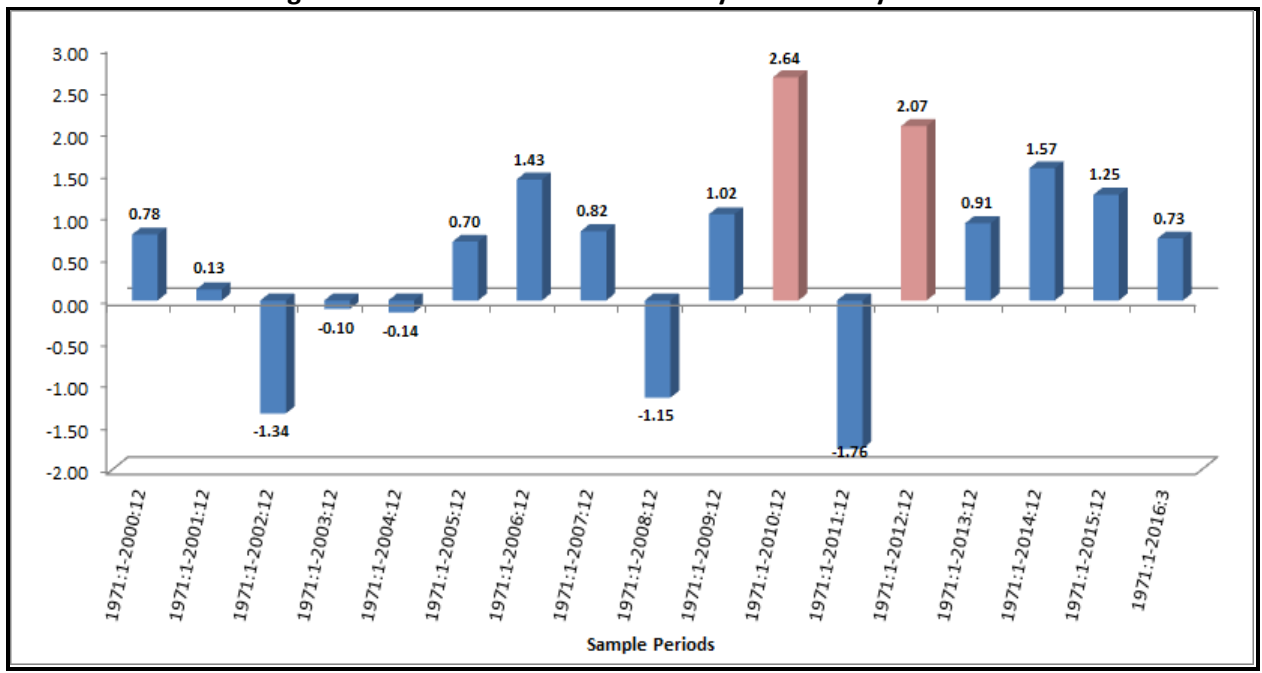

Figure 7: Final State Inflation Elasticity of Monetary Growth

Despite these challenges since 2010, the IT framework produced the longest consecutive (31-months) single-digit inflation episode in the history of Ghana between June 2010 and December 2012. The unprecedented low inflation during the period was largely facilitated by two major disinflation factors. First, considerable foreign exchange inflows from the production and export of crude oil resulted in relative stability in exchange rate in 2010 and 2011 which dampened the inflationary pressures. The second factor is the continuous tighter monetary policy stance (the MPR was increased from 12.5% in December 2011 to 15.0% in June 2012 and was maintained throughout the year) which also countered the high inflation elasticity of money growth observed in 2012. Undoubtedly, the inflation targeting framework has contributed immensely in anchoring economic agents' inflation expectations, leading to better inflation outcomes and inflation variability. The foregoing empirical evidence (i.e. the negative relationship between money supply growth and inflation) clearly lends support to the MPC's decision to adopt the inflation targeting framework.

4.4     Shocks to the System

Figure 8 presents evidence of shocks to the system over the study period. We observe high volatile shocks between 1971 and 1985, a period synonymous with direct control regime and high levels of monetary accommodation of fiscal deficits. No doubt, inflation and inflation volatility were extremely high during this period. Following the erratic pattern of shocks in 1970s and 1980s, we identify two inflationary shocks in 1994 and 2011-12 and three disinflationary shocks in 1991-92, 1995-97, and 2010-11. The economy did not benefit from the disinflationary shocks prior to 2010-11 due perhaps to the absence of a formalized credible framework to help anchor inflation expectations and rein in the entrenched high inflation expectations at the time.

**Figure 8: Shocks to the system**

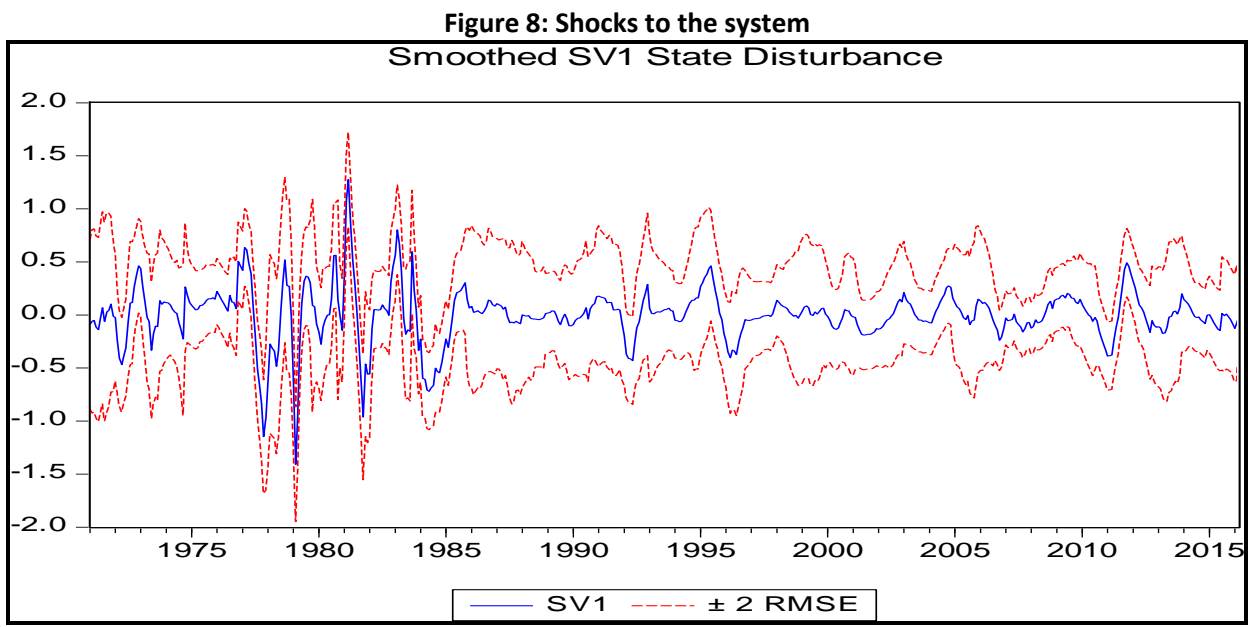

By 2010, however, the IT framework had gained traction amid the official announcement in 2007 that the Bank of Ghana had adopted inflation targeting as a framework of monetary policy. This formalized policy framework facilitated the unprecedented two and half years of single-digit inflation following the disinflationary shock in the 2010-11. Again, the recent significant deviation of inflation from target could be attributed to the inflationary shock of 2011-12.

## 5.     Summary and Concluding Remarks

This paper empirically examines the rationale for the adoption of inflation targeting by Bank of Ghana in 2002. In this regard, it determines the stability or otherwise of the relationship between money supply and inflation in Ghana over the period 1970M1-2016M3. We deplored a battery of econometric techniques, especially CUSUM and recursive coefficients tests for parameter instability, and developed a simple monetarist state-space framework with time-varying coefficients based on the kalman filtration to detect possible changing elasticities of inflation to monetary growth.

Fascinatingly, we generally observed an unstable link between inflation and monetary growth in Ghana, as evidenced by the CUSUM and recursive coefficient tests. In addition, we observed that the final state coefficient is positive but statistically insignificant. This connotes that inflation elasticity of monetary growth was not necessarily explained by monetary growth as at end-March 2016. Consequently, the use of monetary targeting would be an inappropriate framework of monetary policy.

Notably, although the inflation elasticity of monetary growth continued its downward trend into the 1990s and registered several negative readings, it witnessed some appreciably positive and statistically significant values by the middle of the decade. The significant relationship between money supply growth and inflation within this period serves as a clear justification of the adoption of monetary targeting framework by the Bank of Ghana in the 1990s.

However, downward trajectory of the inflation elasticity to monetary growth continued into the 2000s, turning negative between August 2001 and September 2004. Again, the average inflation elasticity for the entire decade (January 2000 to December 2009) was very low compared with the levels recorded in the preceding decades. This clearly suggests that upholding the monetary targeting as a monetary policy framework during the period January 2000 to December 2009 would have yielded very little in terms of delivering the desired inflation outcome.

In lieu of the foregoing evidence of negative relationship between money supply growth and inflation in recent years, the current study lends support for the Bank of Ghana's decision to adopt inflation targeting (IT) framework in 2002. There is no doubt that the IT framework has contributed immensely in anchoring economic agents' inflation expectations, leading to better inflation outcomes and inflation variability. This is because the IT framework produced an unprecedented episode of single-digit inflation from June 2010 to December 2012, reinforced by exchange rate stability (in 2010-11) and tighter monetary policy stance (in 2012) which quelled the observed inflationary shocks in 2010 and 2012. The paper therefore recommends a continuous pursuance and strengthening of the inflation targeting framework in Ghana, as it embodies a more eclectic approach to policy formulation and implementation.

**References**

Andoh S. K., and D., Chappell (2002). "Stability of the money demand function: evidence from Ghana" Applied Economics Letters, 9 (13): 875-878

Ayensu, E. S., (2007). "Bank of Ghana: Commemoration of the Golden Jubilee", Bank of Ghana.

Bawumia, M. (2010). "Monetary Policy and Financial Sector Reform In Africa: Ghana's Experience" Combert Impressions Ghana Ltd, ISBN: 978-1-4507-3403-5

Bawumia, M., Amoah, B., and Z., Mumuni, (2008). "Choice of Monetary Policy Regime in Ghana" Working Paper, Monetary Policy Analysis and Financial Stability Department, WP/BOG-2008/07, Bank of Ghana.

Bawumia, M. and P., Abradu-Otoo, (2003). "Monetary Growth, Exchange Rates and Inflation in Ghana: An Error Correction Analysis" Bank of Ghana Working paper WP/BOG-2003/05, pp. 1-15

Ghartey, E. E. (1998). "Monetary Dynamics in Ghana: Evidence from Cointegration, Error Correction Modelling, and Exogeneity" *Journal of Development Economics*, 57 (2): 473-486.

Heintz, J. and L. Ndikumana, (2010). "Is there a case for formal inflation targeting in sub-Saharan Africa?" Working Papers Series N° 108. African Development Bank, Tunisia.

IMF Institute for Capacity Development/Africa Training Institute (2016). Course on Macroeconomic Forecasting. L-10 State Space Models – Answer Key. Ebène, Mauritius.

Jahan, S. (2012). "Inflation Targeting: Holding the Line" Finance & Development. International Monetary Fund

Kalman, R. E. (1960). "A new approach to linear filtering and prediction problems" Journal of Basic Engineering. 82 (1): 35–45

Kalman, R. E. (1963). *"Mathematical Description of Linear Dynamical Systems"*. Journal of the Society for Industrial and Applied Mathematics Series, 1(2): 152–192.

Kalman, R. E. and R. S. Bucy, (1961). "New results in Linear Filtering and Prediction Theory" Journal of Basic Engineering.

Kovanen, A., (2011). "Monetary Policy Transmission in Ghana: Does the Interest Rate Channel Work?" IMF Working Paper, No. 11/275, pp. 1-33.

Kwakye, J. K., 2012. "Determination of Real Exchange Rate Misalignment for Ghana" Institute of Economic Affairs Monograph, No. 31.

Rummel, O., (2011). "State-Space Models of US (trend) Inflation", Centre for Central Banking Studies, Bank of Ghana

**Appendix A**

1. Sub-Sample Analysis of Final State Elasticity of Inflation to changes in Money Supply

| Sample Periods | Inflation Elasticity of Monetary Growth | P-Values |
|---|---|---|
| 1971:1-2000:12 | 0.78 | 0.484 |
| 1971:1-2001:12 | 0.13 | 0.938 |
| 1971:1-2002:12 | -1.34 | 0.216 |
| 1971:1-2003:12 | -0.10 | 0.958 |
| 1971:1-2004:12 | -0.14 | 0.923 |
| 1971:1-2005:12 | 0.70 | 0.399 |
| 1971:1-2006:12 | 1.43 | 0.350 |
| 1971:1-2007:12 | 0.82 | 0.718 |
| 1971:1-2008:12 | -1.15 | 0.469 |
| 1971:1-2009:12 | 1.02 | 0.411 |
| **1971:1-2010:12** | **2.64** | **0.076*** |
| 1971:1-2011:12 | -1.76 | 0.315 |
| **1971:1-2012:12** | **2.07** | **0.059*** |
| 1971:1-2013:12 | 0.91 | 0.289 |
| 1971:1-2014:12 | 1.57 | 0.397 |
| 1971:1-2015:12 | 1.25 | 0.278 |
| 1971:1-2016:3 | 0.73 | 0.411 |

Note: *** denotes 10% significant level